\documentclass[preprint,aps]{revtex4}

\usepackage{xspace}
\usepackage{graphicx}
\usepackage{dcolumn}
\usepackage{bm}

\def\beq{\begin{equation}}
\def\eeq{\end{equation}}

\def\bea{\begin{eqnarray}}
\def\eea{\end{eqnarray}}

\newcommand{\roughly}[1]%
    {{\mathrel{\raise.3ex\hbox{$#1$\kern-.75em\lower1ex\hbox{$\sim$}}}}}

\newcommand{\gsim}{\mathrel{\roughly>}}

\newcommand{\GeV}{\ensuremath{\mathrm{~GeV}}}
\newcommand{\TeV}{\ensuremath{\mathrm{~TeV}}}

\def\Zp{Z^\prime}
\def\Wp{W^\prime}
\def\Afb{A_{FB}^{t}}

\def\={\,=\,}

\begin{document}

\preprint{CERN-PH-TH/2011-065}
\preprint{MCTP-11-14}

\title{Top quark asymmetry from a non-Abelian horizontal symmetry}

\author{Sunghoon Jung}\email{jungsung@umich.edu}
\author{Aaron Pierce}\email{atpierce@umich.edu}
\affiliation{Michigan Center for Theoretical Physics, Department of Physics,
University of Michigan, Ann Arbor, MI 48109}

\author{James D. Wells}\email{jwells@umich.edu}
\affiliation{Michigan Center for Theoretical Physics, Department of Physics,
University of Michigan, Ann Arbor, MI 48109,}
\affiliation{CERN Theoretical Physics (PH-TH), CH-1211 Geneva 23, Switzerland,}
\affiliation{Cavendish Laboratory, J.J. Thomson Avenue, Cambridge CB3 0HE, UK}


\begin{abstract}
Motivated by the persistence of a large measured top quark forward-backward asymmetry at the Tevatron, we examine a model of non-Abelian flavor gauge symmetry. The exchange of the gauge bosons in the $t$-channel can give a large $A_{FB}^t$ due to the forward Rutherford scattering peak.   We address generic constraints on non-Abelian $t$-channel physics models including flavor diagonal resonances and potentially dangerous contributions to inclusive top pair cross sections. We caution on the general difficulty of comparing theoretical predictions for top quark signals to the existing experimental results due to potentially important acceptance effects.  The first signature at the Large Hadron Collider can be a large inclusive top pair cross section, or like-sign dilepton events, although the latter signal is much smaller than in Abelian models. Deviations of the invariant mass distributions at the LHC will also be promising signatures. A more direct consistency check of the Tevatron asymmetry through the LHC asymmetry is more likely to be relevant at a later stage.
\end{abstract}
\maketitle

\section{Introduction}

Recently the CDF collaboration published an update of its observation of a forward-backward asymmetry in top quark production \cite{Aaltonen:2011kc,cdf:10436}. Focusing on the high-energy region where new physics effects might be expected to be most obvious, CDF measured
$A_{FB}^+	= 0.475 \pm 0.114,$	
where $A_{FB}^{+}$ is the asymmetry of top production in the $t \bar{t}$ rest frame restricted to $m_{t \bar{t}} > 450$ GeV.  For comparison, the SM predicts $A_{FB}^{+}  = 0.088 \pm 0.013$ \cite{Aaltonen:2008hc}.
This measurement builds on previous intriguing inclusive measurements of the forward-backward asymmetry~\cite{cdf:afb,Aaltonen:2008hc,abazov:2007qb}, that consistently yielded large values.
The Standard Model (SM) inclusive prediction \cite{Kuhn:1998jr,Kuhn:1998kw,Bowen:2005ap,Almeida:2008ug} is dominated by ${\mathcal O}(\alpha_{S}^{3})$ QCD interference effects.
 In the $t \bar{t}$ rest frame, CDF measured an inclusive asymmetry (corrected for acceptance)  of $\Afb = 0.158 \pm 0.074$ \cite{cdf:afb}  to be compared with the SM prediction $\Afb = 0.058 \pm 0.009$.  (The inclusive measurement in the lab frame is $\Afb= 0.15 \pm 0.055$  to be compared with a theory prediction $A_{FB}^{t}=0.038 \pm 0.006$ \cite{Bowen:2005ap}.)
Interestingly, two recent separate measurements of   asymmetry in dileptonic top decays were made.  The first made no attempt to reconstruct the top quark system -- it simply measured the asymmetry in the lepton directions and gave a background subtracted result of  $A^{\Delta \eta_{\ell}}_{sub} = 0.21 \pm 0.07$.  The second measurement completely reconstructs the top system and gave a $A_{t \bar{t}}^{\ell \ell} = 0.21 \pm 0.07$.  This results that corresponds to an acceptance-corrected asymmetry  of $A_{FB}^{\ell \ell} = 0.42 \pm 0.16$ to be compared with the SM prediction $A_{FB}^{\ell \ell} = 0.06 \pm 0.01$. In summary, a number of measurements (some completely independent) display an (positive) asymmetry in excess of the Standard Model.  While the results vary in significance, the consistent excess motivates further study.  In this work, we construct a model with a large inclusive rest-frame asymmetry $\Afb \sim 15\%$ and comment on the $m_{t \bar{t}}$ dependence of the asymmetry \footnote{We do not make an effort to fit the other asymmetry measurements.  In any model of new physics, these observations would be strongly correlated.}.

 Previously, we explored the possibility that the exchange of a flavor changing $Z^{\prime}$ in the $t$-channel might give rise to a large asymmetry \cite{Jung:2009jz}.  The model was defined by a non-zero coupling between right-handed up quarks and right-handed top quarks.  That model predicts the  production of like-sign top quarks.  Indeed, the non-observation of like-sign tops at the Tevatron already placed strong constraints on the model.  Owing to the large $uu$ parton luminosity at the LHC, one would expect a strong signal there soon in such models, if they are not already excluded.  This signal was recently studied by \cite{Berger:2011ua,Gupta:2010wx,Cao:2011ew}, and although the scenario of~\cite{Jung:2009jz} may not be ruled out by direct analyses, its parameter space appears to be severely constrained (if not ruled out) by reinterpretations of other LHC results.

Here we build a model based on $SU(2)_X$ flavor symmetry, first discussed in \cite{Jung:2009jz},  that places the $(u \;  t)_R$ together in a doublet \footnote{In principle, one could extend this structure to a full SU(3) that would include the charm quark.   However, this does not help with the explanation of the $\Afb$, and depending upon implementation could actually give rise to dangerous contributions to $D$--$\bar{D}$ mixing.}.
This model explains $\Afb$ measurement dominantly by the exchange of the ``$\Wp$" bosons of the theory that raise and lower the quarks in this doublet.  The non-Abelian nature of the flavor symmetry ensures that the gauge bosons carry ``top number," just as gluons carry color.  The conservation of top number prevents the production of like-sign tops via $uu \rightarrow tt$ or $u g \rightarrow t \Wp \rightarrow t t u$, which are present in the Abelian model \cite{Jung:2009jz}. However, we will argue that a small breaking of top number is both allowed and has the benefit of opening new parameter space for this model. The result is a small, but potentially observable like-sign top signal.

By analyzing this example model, we also hope to address various general features of (non-Abelian) $t$-channel physics (see also \cite{Cheung:2009ch, Shu:2009xf, Arhrib:2009hu, Dorsner:2009mq, Jung:2009pi, Barger:2010mw, Cao:2010zb, Xiao:2010hm, Jung:2010yn, Cheung:2011qa, Shelton:2011hq, Bhattacherjee:2011nr, Barger:2011ih, Gresham:2011dg, Patel:2011eh, Grinstein:2011yv, Barreto:2011au, Ligeti:2011vt, AguilarSaavedra:2011vw, Gresham:2011pa}). It is by now well recognized that, despite the ease at which it generates a large asymmetry, $t$-channel physics is constrained by large like-sign top pair production and an enhancement of top pairs in the high $m_{t\bar{t}}$ region. We revisit these issues both at the Tevatron and LHC7, including both flavor changing and conserving couplings of the $\Wp$, and consider potentially important contributions to inclusive top pair sample from the production of the new gauge bosons associated with the top quark. In addition, we study important constraints coming from the $\Zp$, the mostly flavor conserving resonance particle accompanied by the $\Wp$ in any non-Abelian model of this general class.

Another important aspect of $t$-channel physics that we discuss in detail is acceptance effects. The very forward top quarks abundantly produced by the Rutherford scattering peak are not identified as easily as more central top quarks. As a result, care is needed to interpret the data and determine what theories are viable.

\section{A model}

To the SM, we add a $SU(2)_X$ gauge symmetry.  The $(u \; t)_R$ is the only SM field charged under this symmetry.  This matter content is anomalous (including the global Witten anomaly \cite{Witten:1982fp}).  A UV complete theory would involve the introduction of new states charged under both $SU(2)_{X}$ and $SU(3)_C\times U(1)_Y$.  It is plausible that these states would be near the TeV scale, and could thus be produced at the LHC.  We will not concern ourselves with the details of the UV completion, but concentrate on the  consequences of the gauge boson interactions with SM fermions.  Also, we do not discuss flavor constraints, but merely state here that they can be tamed to acceptable levels by tuning the Yukawa couplings appropriately.

The gauge bosons masses in the $SU(2)_{X}$ sector depend on the choice of Higgs boson representations.  A single Higgs doublet leaves a custodial symmetry intact, leading to a degenerate  $\Wp$ and $\Zp$.    (We emphasize that the $\Wp$ bosons do not carry an electric charge -- the notation merely indicates that they are responsible for raising and lowering within the $SU(2)_X$ multiplet.) For a Higgs multiplet with $X$-isospin $T$, we have
\begin{equation}
M_{\Wp}^2 = g_X^2 v_T^2 \frac{ T(T+1)-T_3^2}{2} , \qquad M_{\Zp}^2 = g_X^2 v_T^2 T_3^2,
\end{equation}
where we have assumed it is the $T_{3}$ component that gets a vacuum expectation value.  For phenomenological reasons described below, it is advantageous to consider the possibility of gauge bosons that are not precisely degenerate.  In our studies, we consider both $M_{\Wp}$ and $M_{\Zp}$ as free parameters.  In many cases, this can be accomplished by including two Higgs bosons in different representations that contribute to the masses. However, the maximum size of the hierarchy between the two masses, $M_{\Zp}/M_{\Wp}$, is limited by the size of the Higgs representation (see also \cite{Barger:2011ih}).  It is for this reason that we do not allow a too heavy $\Zp$ gauge boson. Consequently, its resonant production will be important and discussed in the next section.

The presence of non-degenerate gauge boson masses allows the possibility of a physical rotation corresponding to a mismatch between the $(u \; t)_R$ gauge eigenstates and the mass eigenstates.  We parameterize this mismatch by an angle $\theta$. With this definition, we have the fermion interaction Lagrangian
\begin{eqnarray}
{\cal L} &=& \frac{g_X}{\sqrt{2}} W'^{-}_\mu \Big\{ \bar{t}_R \gamma^\mu t_R (-cs) \,+\, \bar{u}_R \gamma^\mu u_R (cs) \,+\, \bar{t}_R \gamma^\mu u_R (c^2) \,+\, \bar{u}_R \gamma^\mu t_R (-s^2) \Big \} \nonumber\\
&& +\, \frac{g_X}{\sqrt{2}} W'^{+}_\mu \Big\{ \bar{t}_R \gamma^\mu t_R (-cs) \,+\, \bar{u}_R \gamma^\mu u_R (cs) \,+\, \bar{t}_R \gamma^\mu u_R (-s^2) \,+\, \bar{u}_R \gamma^\mu t_R (c^2) \Big \} \nonumber\\
&& +\, \frac{g_X}{2} \Zp_\mu \Big\{ \bar{t}_R \gamma^\mu t_R (c^2-s^2) \,+\, \bar{u}_R \gamma^\mu u_R (s^2-c^2) \,+\, \bar{t}_R \gamma^\mu u_R (2cs) \,+\, \bar{u}_R \gamma^\mu t_R (2cs) \Big\}.
\label{interaction}
\end{eqnarray}
where $c=\cos \theta$ and $s=\sin \theta$. Top number is broken by nonzero $\theta$. If the $\theta$ becomes too large, dangerously large like-sign top quark production will re-emerge.   Also, additional bounds from diagonal $u\bar u$ resonant production of the $\Wp$ also appear for $\theta \neq 0$.  For both of these reasons, we expect $\cos \theta$ should be close to 1. However, $\cos \theta \ne 1$ does play an important role as will be more quantitatively discussed in Sec. \ref{sec:dijet} and \ref{sec:TevMeas}.

Presumably, the presence of this new $SU(2)$ symmetry would be linked with the flavor puzzle of the SM.  We will not attempt to build a full theory of flavor.  We content ourselves to note that if one of the Higgs bosons that breaks the $SU(2)_X$ is a doublet, then effective SM Yukawa couplings can  arise from $d=5$ operators.
\begin{eqnarray}
\Delta {\cal L} &\ni&  \frac{(\lambda_u^\prime)_i}{M} \,( \bar{Q}_i \cdot h_{SM} )\, (\phi_D \cdot q).
\end{eqnarray}
Here, $i=1,2,3$ is a generation index, and the contractions of $SU(2)_{L,X}$ indices are denoted by dots. $Q$ is SM left-handed doublet while $q=(t_R,u_R)$ is $SU(2)_X$ doublet. $M$ can be thought of as a mass scale a flavor sector.
Because the top quark Yukawa arises from this higher dimension operator,  the $v_D/M$ ratio cannot be too small.  Thus, the new colored states associated with the scale $M$ are likely accessible at the LHC.  Furthermore, this fact argues that there should be a large doublet vev;
other vevs would not allow low dimension Yukawa couplings.  This places a restriction on the hierarchy between the gauge boson masses.  All benchmarks that we present can achieve the necessary top quark mass with $ \lambda_u^\prime$ perturbative and with only a doublet and triplet Higgs boson of the $SU(2)_X$.


In Table \ref{tab:points}, we present benchmark points whose phenomenology we will analyze in some detail.  The points are chosen to give substantial forward-backward asymmetries ($\sim15\%$), while giving rough agreement with the total top production cross section.  They are all consistent with the important dijet constraints outlined below.  Upon detailed examination, point $A$ appears consistent with all known data, and represents a ``best point" for this model.  Points $B$ and $C$ are discussed in part to illustrate what happens when one deviates from this ``best region.''  Point $B$ has tension with the observed measurements of the top quark production rate. Point $C$ appears to be excluded by measurements of the invariant mass distribution in top quark production.


\begin{table}
\begin{tabular}{@{\hspace{0.3cm}} c @{\hspace{0.7cm}} c @{\hspace{0.5cm}} c @{\hspace{0.5cm}} c @{\hspace{0.5cm}} c}
\hline
\hline
& $M_{W'} \, (\GeV)$ & $M_{Z'}  \,  (\GeV) $ & $\alpha_{X}$  & $\cos \theta$\\
\hline
A: & 200 & 280 & 0.060 & 0.95 \\
B: & 200 & 80 & 0.044 & 0.95 \\
C: & 850& 1200& 0.75 &  1 \\
\hline \hline
\end{tabular}
\caption{Benchmark points to be explored below. Point $A$ represents a best point consistent with all data.  Points $B$ and $C$ are in tension with measurements on top production from the Tevatron as outlined in Sec. \ref{sec:TevMeas}.  Points are selected to give appreciable $\Afb$ while avoiding constraints from dijet searches (see Sec. \ref{sec:dijet}), and maintaining a rough agreement with the total top production cross section. $\alpha_X \equiv g_X^2/4\pi$.} \label{tab:points} \end{table}

\section{Constraints from Dijet Measurements}\label{sec:dijet}

Before turning to a detailed discussion of the prediction of the forward-backward asymmetry, we consider the bounds on this class of models arising from dijet events.
While $t$-channel exchange of the $\Wp$ is responsible for the bulk of the $\Afb$ in these models, very important phenomenological constraints arise from the (approximately flavor conserving)  $Z^{\prime}$.  As noted above, it is theoretically possible to decouple the $Z^{\prime}$ through an appropriate choice of the representations of the Higgs boson.  However,  a large hierarchy would be needed to completely decouple the effects of the $Z^{\prime}$, requiring uncomfortably large Higgs representations.  Therefore, we take some care to treat the constraints arising from a $Z^{\prime}$ boson that is relatively light and not completely decoupled. Note, the $\Zp$ is narrow, as discussed in the appendix~\ref{app:xsec}. Thus resonance searches are meaningful.

\begin{figure}
\includegraphics[width=0.64\textwidth]{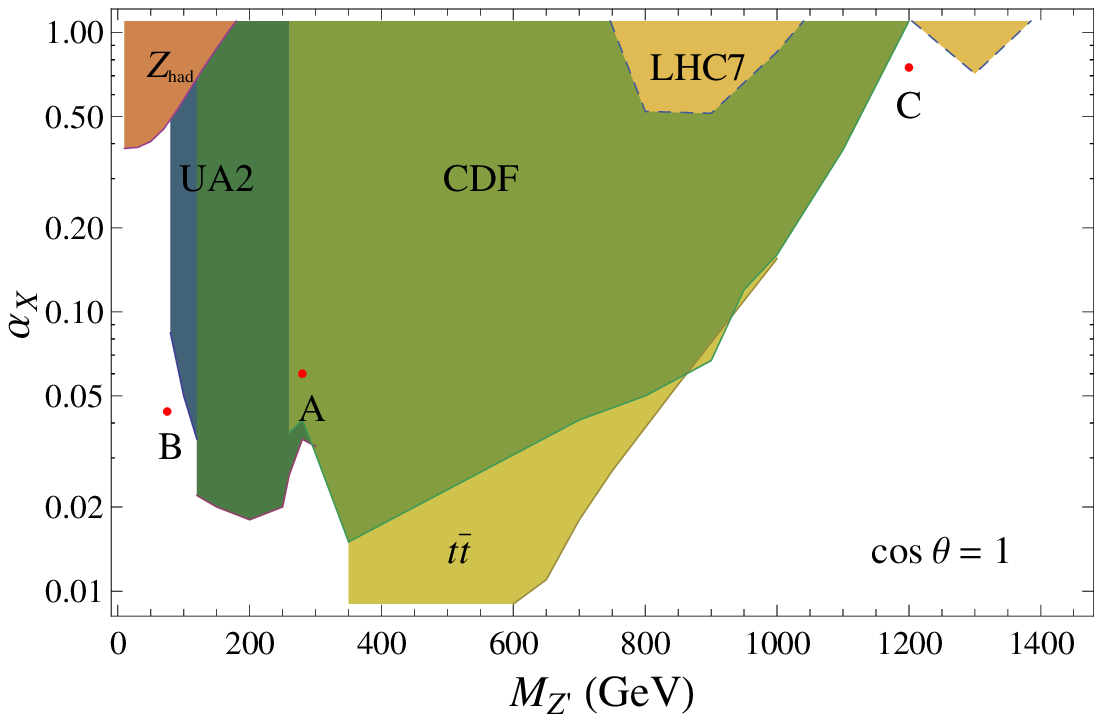}
\includegraphics[width=0.64\textwidth]{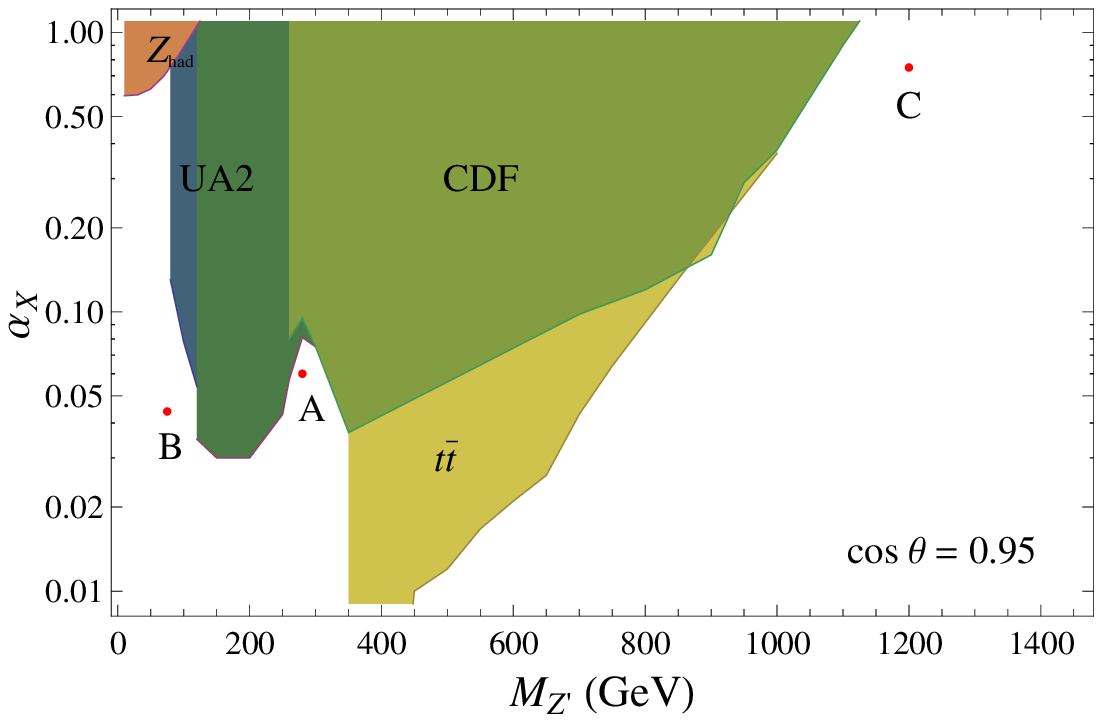}
\caption{Bounds in the $\{ M_{Z'}, \alpha_{X} \}$ plane.  Exclusion limits are obtained by considering constraints arising from one-loop corrections to the hadronic width of the $Z$ boson \cite{Nakamura:2010zzi,Carone:1994aa}, searches for dijet resonances at UA2 and CDF \cite{Alitti:1990kw,Alitti:1993pn,Aaltonen:2008dn} (UA2 results from the first and second stage running are shown in separate colors), angular distribution of dijet events at the $7 \TeV$ LHC \cite{Collaboration:2010eza} and the combined $t \bar{t}$ resonance searches at CDF and D0 using up to 1 fb$^{-1}$ of data \cite{CDF:2007dia,Abazov:2008ny}.  Also shown are locations of benchmark points $A,B,C$ that will be studied in more detail later.  Plots are shown for $\cos \theta =1$ (top panel) and $\cos \theta =0.95$ (bottom panel).}
\label{fig:dijetbounds}
\end{figure}

The constraints are summarized in Fig.~\ref{fig:dijetbounds}. At low masses, the constraint arising from the one-loop correction to the hadronic width of the $Z$ boson \cite{Nakamura:2010zzi,Carone:1994aa} is important. As masses increase, limits from dijet searches at UA2 \cite{Alitti:1990kw,Alitti:1993pn}, CDF \cite{Aaltonen:2008dn} become important, and finally, LHC7 \cite{Collaboration:2010eza} results at higher invariant mass energies become important.  We have also shown the region excluded by resonant searches in the $t \bar{t}$ production channel \cite{CDF:2007dia,Abazov:2008ny}. We refer only to the most recent published results analyzing up to 1 fb$^{-1}$ of data. Consideration of data on the differential $t\bar{t}$ cross section may somewhat extend the excluded regions shown. We have shown constraints for $\cos \theta =1$ (top panel) and $\cos \theta = 0.95$ (lower panel). We note that the inclusion of non-zero $\theta$ opens up additional regions of parameter space, allowing point $A$ to be consistent with all data. This is due to reduced flavor diagonal couplings of the $Z^{\prime}$.  For all our benchmark points we have $M_{W'} \neq M_{Z'}$.  In this case, we emphasize there is no a priori expectation that $\cos \theta =1$.  In fact, $\theta \rightarrow 0$ might even be viewed as a tuning.

In Fig.~\ref{fig:dijetbounds}, only the constraints from the $Z'$ are included.  In principle, if the devation of $\cos \theta$ from 1 is too large, the resulting flavor {\it conserving} couplings of the $\Wp$ can allow it to be resonantly produced at a dangerous rate.  We find $\cos \theta \gsim 0.92$ is safe for a light $\Wp$ (i.e. with mass below the $t \bar{t}$ threshold).  This  includes Points $A$ and $B$.  For heavier  $W^{\prime}$ masses, the search for a resonance in $t \bar{t}$ represents a stronger constraint.  The $W^{\prime}$ constraint from $u\bar u$ and $t\bar t$ resonance production and decay is satisfied for point $C$ for $\cos \theta \gsim 0.97$.

\section{Consistency with existing Top Cross Section Measurements}\label{sec:TevMeas}
We now address the question of whether our benchmarks are consistent with the detailed cross section measurements of the top quark at the Tevatron.

\begin{figure}
\includegraphics[width=0.7\textwidth]{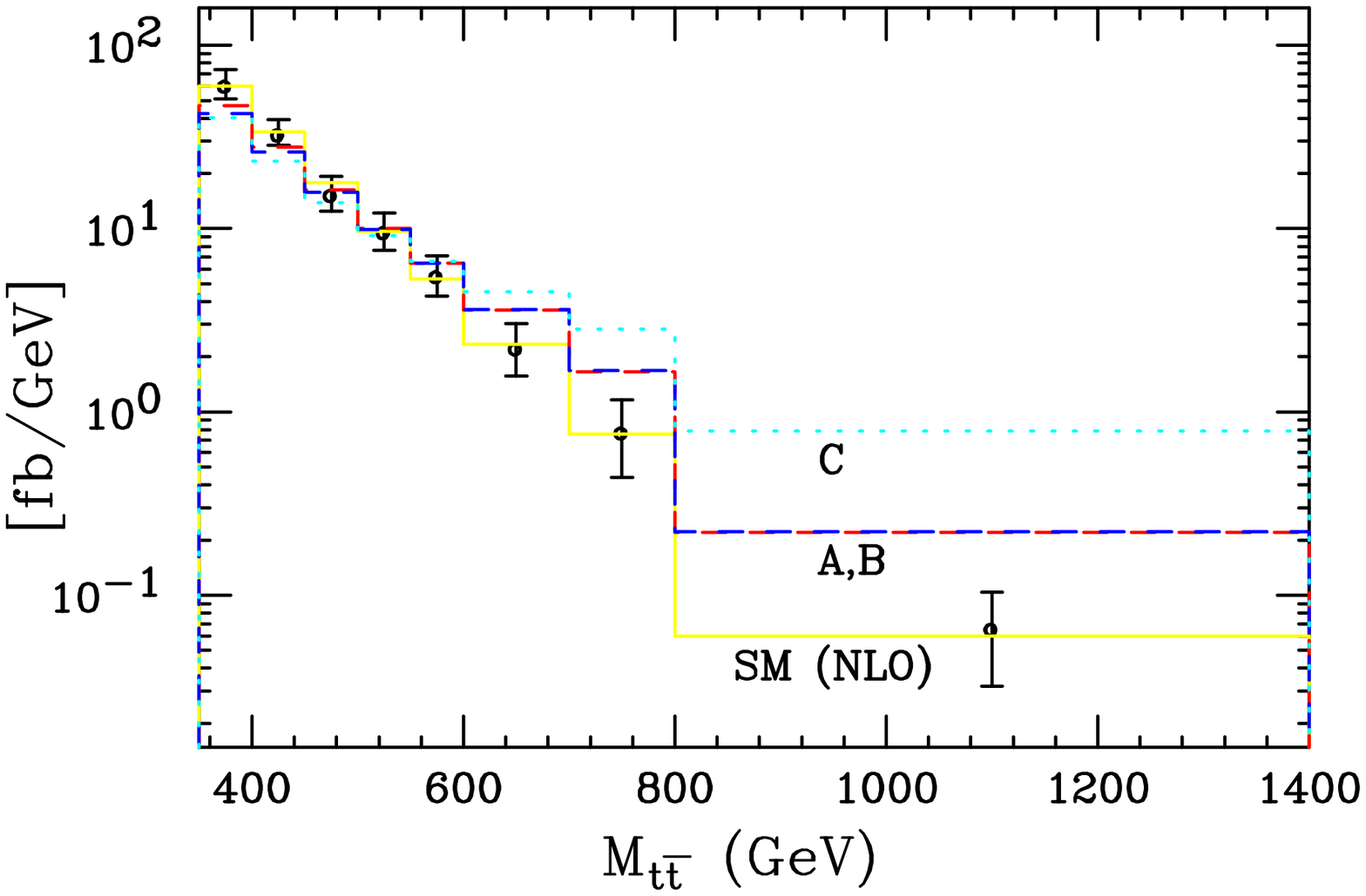}\\
\includegraphics[width=0.3\textwidth]{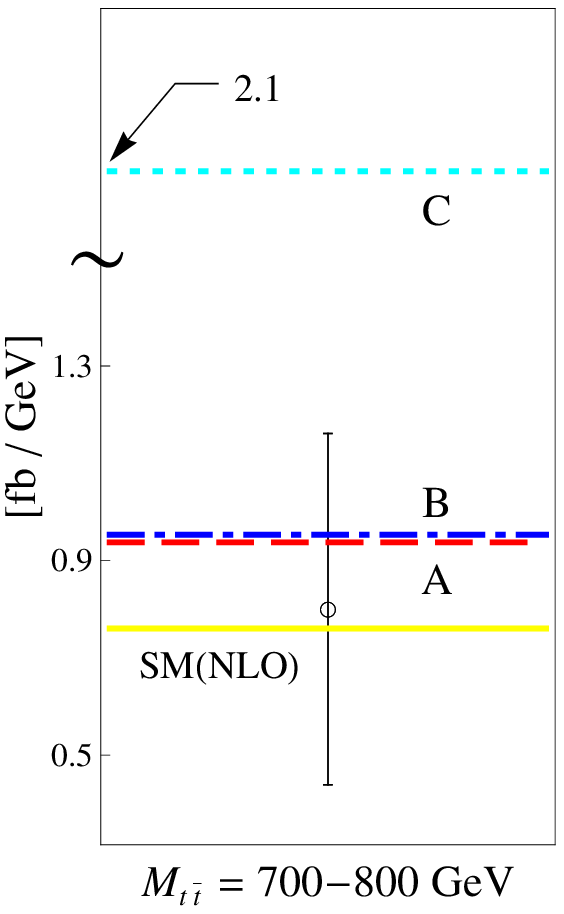}
\includegraphics[width=0.3\textwidth]{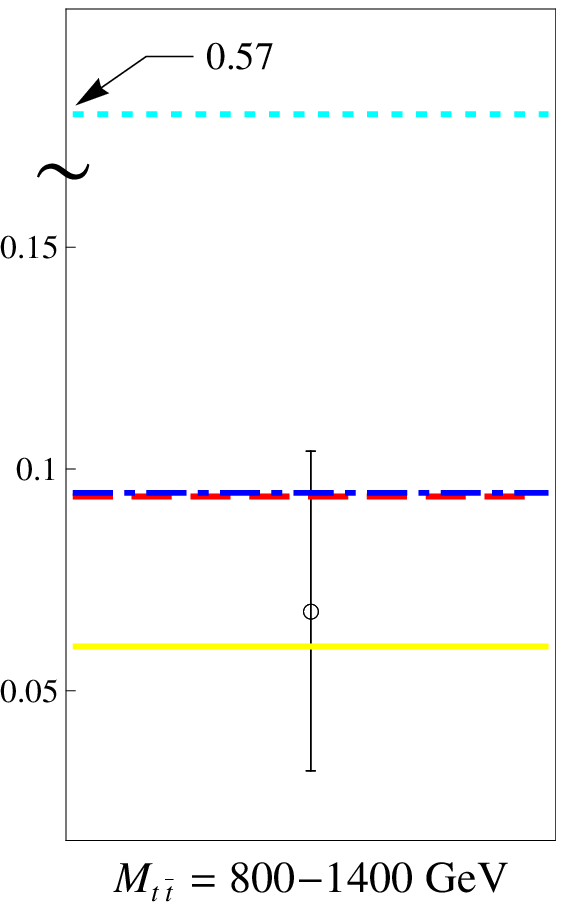}
\caption{The $m_{t\bar{t}}$ distribution is shown for the SM, as well as the three benchmark points $A,B,C$ displayed in Table A.  In the top panel, we show theoretical distributions, after applying a $\hat{s}$ dependent SM NLO $K$-factors  \cite{Nason:1987xz} to all models.  No acceptance cuts are applied. We note a large discrepancy from the measured CDF data \cite{Aaltonen:2009iz}. However, this parton level calculation ignores potentially very important acceptance effects as discussed in the text.  A rough correction for these effects yields the plots in the lower panels for the highest two bins. Points $A$ and $B$ now seem plausibly consistent with the data, whereas point $C$ is still clearly excluded. Faking contributions in the lower panels are about $5\%$ of the true $t\bar{t}$ process for $A$ and $B$.}
\label{fig:mtt}
\end{figure}

An important constraint on these models comes from the $t \bar{t}$ invariant mass distribution \cite{Aaltonen:2009iz}.  With respect to the SM, these models overproduce top quarks at large invariant mass due to the Rutherford enhancement. In Fig.~\ref{fig:mtt}, we show the invariant mass distribution of the $t \bar{t}$ for the benchmark points shown in Table~\ref{tab:points}. We have applied the $\hat{s}$ dependent NLO $K$-factor of the SM \cite{Nason:1987xz} to all distributions (including those with new physics). Absent a proper NLO calculation in these models, this approach represents an optimistic attempt to capture some of the leading QCD corrections. We have used CTEQ6L \cite{Pumplin:2002vw} and CTEQ6.6M \cite{Nadolsky:2008zw} parton distribution sets for the LO and NLO cross sections, respectively. $m_t=172.5$ GeV and $\mu = m_t$ are assumed. A naive examination of the highest $\hat{s}$  bins of distributions shown there would indicate that the new physics models are excluded.

However, this model produces very forward top quarks.  The acceptance for these top quarks is far from assured, and indeed, can be substantially lower than the SM.  The angular behavior deviates most substantially from the SM at the highest partonic center of mass $\sqrt{\hat{s}}$ where the forward scattering peak is most pronounced.   We now investigate whether the large enhancement at high $\sqrt{\hat{s}}$ persists after acceptance effects are addressed.

We model losses of very forward top quarks by modeling the unfolding procedure of the experiments in an approximate but well-defined way. We first generate a parton-level Monte Carlo event sample of the SM in MadGraph/MadEvent v.4.4.492 \cite{Alwall:2007st}, and weight it by an $\hat{s}$-dependent SM NLO $K$-factor.  We take this sample, apply the selection cuts of the CDF $m_{t \bar{t}}$ analysis \cite{Aaltonen:2009iz} and calculate $m_{t\bar{t}}$ using only the leading four jets, a charged lepton and the missing energy as done by CDF. The resulting $d \sigma/d m_{t \bar{t}}$ distribution is compared to the original theoretical distribution prior to the selection cuts.  This comparison allows us to derive a ``smearing matrix" in the binned $m_{t\bar{t}}$ space that estimates how the cuts and reconstruction take a theoretical distribution to a measured one.   We then use this same matrix for all model samples. This includes our benchmark points and generalized color-octet models having $A_{FB}^t = 0.1,\,0.2$ (which are sometimes used to test the experimental unfolding procedure).   Application of the cuts, $K$-factor and, subsequently, the smearing matrix (as derived from the SM distribution) leave the $m_{t\bar{t}}$ distributions of the color-octet models nearly unchanged -- an indication that their acceptance is similar to the Standard Model. This is not the case  for our benchmark points.  Many of the events in the highest $m_{t \bar{t}}$ bins are lost due to the selection cuts.  The result of the above procedure (cuts, $K$-factor, smearing) is shown in the lower panels in Fig.~\ref{fig:mtt}.  As a result, for points $A$ and $B$, the agreement with the data is now quite good.  This illustrates care is needed to account for acceptances when analyzing the viability of these models.  Even with the corrections, point $C$ is excluded, and we do not consider it further.  It indicates that a model with large mass $\Wp$ and $Z^{\prime}$ will have difficulty reproducing this distribution.

A measurement of the total top pair production cross section is, on the other hand, most sensitive to the low $\sqrt{\hat{s}}$ region where the bulk of the parton luminosity lies.  In this region, the interference of $t$-channel $\Wp$ exchange and the SM (which comes with negative sign) dominates over new physics squared contributions. The result is a prediction for a total cross section smaller than the SM value for model points having asymmetries of the size observed at CDF.   Measurements of the total top cross section constrain Model $B$ more than Model $A$.  This is clear from examining the more pronounced suppression  for Model B in the lowest $m_{t\bar{t}}$ bin  in Fig. \ref{fig:mtt}.  This  makes Model $B$ less appealing than $A$.  The cross section of $B$ could be increased by increasing the $\alpha_{X}$, but this would result in an asymmetry that is, at present, too large.  Depending on the evolution of the asymmetry measurements, it may be useful to revisit this region of parameter space.  For now, however, we focus on point $A$.

There is another potential (fake) contribution to inclusive top cross sections coming from final states where the gauge bosons $\Wp, \Zp$ are either pair produced, or produced in association with the top quark, e.g.  $u\bar{u} \to W^{\prime +} W^{\prime -}$ and $gu \to W^{\prime -} t$. These events eventually pollute the inclusive top quark sample after subsequent decays of the $W'^\pm$ bosons.
In general, this pollution will increase the size of the inferred cross sections for each decay channel of top pairs, resulting in a somewhat improved agreement with experiment than the original calculation reported above.  We discuss these effects and their importance more quantitatively in next few paragraphs.

In this $t$-channel model, as in \cite{Jung:2009jz}, the apparent top production cross section will differ in different decay modes. That is, the cross section inferred from the dilepton channel will differ from the semi-leptonic mode. This is largely because the associated productions of $\Wp, \Zp$ mentioned above contribute preferentially to the semi-leptonic mode over the dilepton mode. As shown in table \ref{tab:crxTev}, for point $A$, our parton level results imply that a theoretical $\sigma(t\bar{t}) = 6.7$ pb will be interpreted as $7.0$ pb and $6.6$ pb in the semi-leptonic and dileptonic channels, respectively, once the ``fake" contributions are taken into account.   These contributions are about $13 \%$ and $8\%$ of the true $t\bar{t}$ process, making a small but potentially noticeable difference between the two channels.  For these faking contributions, we have not applied a $K$-factor (which presumably would be very different from the one relevant for $t \bar{t}$.  If one were to apply a $K$-factor of similar size, the results would become  $7.2$ pb and $6.7$ pb for the semi-leptonic and dileptonic modes, respectively.

\begin{table} \centering \begin{tabular}{c @{\hspace{0.5cm}} c @{\hspace{0.3cm}} c @{\hspace{0.3cm}} c @{\hspace{0.3cm}} c}
\hline
\hline
 & $\sigma(t\bar{t})_{thy}$ & $A_{FB}^t$ & $\sigma(t\bar{t})_{lj}$ & $\sigma(t\bar{t})_{ll}$ \\
\hline
CDF & $7.50 \pm 0.48$ pb \cite{cdf:9913} & $0.158 \pm 0.074$ \cite{cdf:afb}& $7.22 \pm 0.79 $ pb \cite{Aaltonen:2010ic} & $7.25 \pm 0.92$ pb \cite{cdf:10163, cdf:9890}\\
SM & 7.34 pb & $0.058 \pm 0.009$ & 7.34 pb (normalized) & 7.34 pb (normalized) \\
Model A & 6.69 pb & 0.14 & 7.0 pb & 6.6 pb \\
\hline \hline
\end{tabular} \caption{Cross sections for point $A$ at the Tevatron. The theoretical value of exclusive top pair cross section is shown in the column denoted $\sigma(t\bar{t})_{thy}$, where we apply a $K$-factor found by normalizing the LO SM to the approximated NNLO results $7.34$ pb, averaged over three independent results \cite{Langenfeld:2009wd, Cacciari:2008zb, Kidonakis:2008mu}. The rest-frame asymmetry $A_{FB}^t$ is also shown. Inclusive top pair cross sections in the semi-leptonic ($\sigma(t\bar{t})_{lj}$) and dileptonic ($\sigma(t\bar{t})_{ll})$ are obtained by applying CDF selection cuts \cite{Aaltonen:2010ic, cdf:10163, cdf:9890} and by including other faking contributions, see text for detail. } \label{tab:crxTev} \end{table}

The fake top cross sections for point A are somewhat suppressed by the choice of $\cos \theta = 0.95$.   This is because for $\cos \theta < 1$ a new decay mode $\Wp \to u\bar{u}$ opens up which does not involve any top quarks.   If instead, $\cos \theta =1$, the top cross sections would be measured as $\sigma(t\bar{t})_{lj,ll} = 8.0, 7.5$ pb with a (somewhat smaller) asymmetry,  $A_{FB}^t = 0.12$.  The decrease in the asymmetry is largely a result of  contributions from the process $gu \to W^{\prime -} t$.    In this $t$-channel process,  the top quark is preferentially produced along the initial gluon direction which is likely from anti-proton,  hence backward.   This somewhat large inferred cross section, coupled with earlier dijet constraints argues for $\cos \theta < 1$.  Furthermore, it is likely that for $\cos \theta = 1$  faking contributions might have already been observed at the LHC (see Sec.~\ref{sec:LHC}).  For values of $\cos \theta$ near, but not equal to one, these events can provide a good LHC signature.

Since points $B$ and $C$ had difficulty reproducing the top quark measurements at the Tevatron, we will confine our detailed analysis of the asymmetry to point $A$.  However, the signatures we point out in the final section are qualitatively applicable to all $SU(2)_{X}$ models.

\section{Phenomenology of the Forward-Backward Asymmetry}

 The same acceptance effect discussed in the previous section can also impact the $A_{FB}$ asymmetry.  Events that would have otherwise contributed to $A_{FB}$ go undetected.  While CDF attempts to correct for this effect (see for example refs.\cite{cdf:afb, Aaltonen:2011kc}), the correction is model-dependent. We have checked that the acceptance for our model can differ significantly from the acceptance of the SM \footnote{We checked that generalized color-octet models having $\Afb=10, 20 \%$ are better corrected under the same correction matrix. We comment that generalized color-octet models were somewhat used by CDF groups to test their unfolding procedure \cite{Aaltonen:2011kc}.}. This results in a dilution of the $A_{FB}$ relative to the theoretical prediction.  This effect can be substantial, again, particularly at the highest $m_{t\bar{t}}$. At the parton level, we predict $\Afb=0.15$, and $A_{FB}^{+} = 0.30$ in the rest frame for point $A$.  However, applying acceptance cuts, and the correction matrix of the size applied at CDF, we estimate a $A_{FB}^{+} = 0.22$ would be observed. Moreover, the growth at the very largest $m_{t \bar{t}}$ (present at the parton level before acceptance effects) would be partially suppressed. This is of interest as the final bin of the CDF measured $\Afb$ in $m_{t \bar{t}}$ actually shows a decrease \cite{Aaltonen:2011kc}, although with large error bars.  So, while this model does not predict a decrease as observed, due to acceptance effects, a rapid rise is not observed either.

The procedure used to estimate the above results is similar to the method presented in Sec. \ref{sec:TevMeas}. In this case, the correction matrix for the asymmetry is in the four-binned $-Q_l \cdot \eta(t)$ space as used by the CDF collaboration \cite{cdf:afb}.

Although it is a proton--proton collider, the LHC can also measure the forward-backward asymmetry of the top quark. A reference direction can be provided by the boost direction of the top pair \cite{Jung:2009jz,Wang:2010tg}. In a $q\bar{q} \to t\bar{t}$ subprocess, the initial valence quark $q$ is likely to be more energetic than the initial sea quark $\bar{q}$.  A forward top quark inherits this boost.  Thus we define an asymmetry with respect to the boost direction as
\beq
A_{boost} \= \frac{ N( a>0) - N( a<0) }{ N(a>0) + N(a<0) }, \qquad a \equiv (y_t + y_{\bar{t}})(y_t - y_{\bar{t}}).
\label{eq:afbboost} \eeq
where top rapidity is denoted by $y_t$.   Note the $(y_{t} + y_{\bar{t}})$ factor in $a$ measures the direction of the boost of the $t\bar{t}$ system, while the  $(y_{t} - y_{\bar{t}})$ gives the direction of the asymmetry.  Again, by restricting to the high energy region, one can measure a higher asymmetry. This also tends to suppress the symmetric $gg\to t\bar{t}$ subprocess. We estimate $A_{boost} \cong 0.06$ at LHC7 for point $A$ with $m_{t\bar{t}} \geq 450$ GeV. It is smaller than the Tevatron values because the $q\bar{q}$-initiated process is less important, and there is $\sim 25\%$ chance of a mismatch between the boost and $q$ directions. Another strategy is to focus on the central region where the $gg$-initiated process is relatively small \cite{Ferrario:2008wm}.  While the observable, $A_{boost}$, is unlikely to  be the optimal LHC discovery mode of the model --  a large amount of data would be required to reconstruct the observable -- it might provide a more direct consistency check of the Tevatron asymmetry measurement (see also \cite{Craig:2011an}). New physics indications will come more quickly at the LHC by other observables, as we discuss in the next section.

\section{LHC and Discussion} \label{sec:LHC}

At the LHC, one big difference from the Tevatron is that the gluon luminosity is much larger. Consequently, our new physics effects on exclusive top pair production $pp \to t\bar{t}$ (that rely on a  $q\bar{q}$ initial state) is relatively small. As shown in table \ref{tab:LHC}, the total cross section $\sigma(t\bar{t})_{thy}$ for Point A differs from the SM only by a small quantity, unlike at the Tevatron.

However, the inclusive cross section can be significantly affected by gluon-initiated associated production of gauge bosons. $gu \to \Wp t$ is the most important, as its cross section becomes similar to honest top pair production \cite{Shu:2009xf, Dorsner:2009mq, Barger:2011ih, Cheung:2011qa}. Since LHC7 has already performed rough measurements of the cross sections (see Table \ref{tab:LHC}), the large inclusive cross section predicted by $t$-channel models are potentially already constrained. By applying selection cuts from ATLAS analysis \cite{Aad:2010ey} and by including all processes contributing, $t \bar{t},\, tt,\, \bar{t} \bar{t},\, t V, \bar{t} V$ and $V V$ (where V = $\Wp$, $\Zp$), we estimate inclusive cross sections in both semi-leptonic and dileptonic channels $\sigma(t\bar{t})_{lj,\,ll}$.
We apply a $K$-factor $K = 1.89$ (appropriate for normalizing LO SM to the approximated NNLO) to every diagram with $t\bar{t}$ exclusive final states. Associated production of gauge bosons are calculated at LO. Faking contributions are dominantly from the process $\sigma(gu \to \Wp t) = 47$ pb (LO) for point A that contributes $14\%$ and $12\%$ of true $t\bar{t}$ events in two channels, respectively \footnote{If we had multiplied by the same $K$-factor for faking contributions (though there is no reason to expect this is the correct value), the predictions for the observed inclusive cross-sections would become  $\sigma(t\bar{t})_{\ell j} = 215$ pb and  $\sigma(t\bar{t})_{\ell \ell} = 195$ pb.}. See table \ref{tab:LHC}.

\begin{table}
\begin{tabular}{c @{\hspace{1cm}} c @{\hspace{0.5cm}} c @{\hspace{0.5cm}} c @{\hspace{0.5cm}} c}
\hline
\hline
& $\sigma(t\bar{t})_{\ell j}$ & $\sigma(t\bar{t})_{\ell \ell}$ & $\sigma(t\bar{t})_{thy}$ & $ A_{boost}$ \\
\hline
ATLAS \cite{Aad:2010ey} & $142^{+61}_{-46}$ pb & $151^{+86}_{-66}$ pb  & $145^{+52}_{-41}$ pb  & ---\\
CMS   \cite{Khachatryan:2010ez} & ---  &  $194_{\pm 79}$ pb & --- & --- \\
Model A &  193 pb &  177 pb & 166 pb & $6\%$ \\
\hline \hline
\end{tabular}
\caption{Detailed LHC7 cross section predictions for Point $A$. Exclusive $pp \to t\bar{t}$ cross section is shown in the column denoted by $\sigma(t\bar{t})_{thy}$, with a $K=1.89$ normalizing LO SM to the approximate NNLO SM calculation $\sigma(t\bar{t}) = 164.6^{+11.4}_{-15.7}$ pb \cite{Aad:2010ey, Moch:2008qy, Kidonakis:2010bb}.  The other two columns $\sigma(t\bar{t})_{\ell j}$ and $\sigma(t\bar{t}) _{\ell \ell}$ represent predictions for observed inclusive cross-sections in the semi-leptonic and dileptonic channels. Here we have included possible ``fake" contributions dominantly from the process $\sigma(gu \to \Wp t) = 47$ pb at the leading order and applied the cuts from the ATLAS analysis \cite{Aad:2010ey}. See text for more discussions. $A_{boost}$ observable is defined in Eq.~\ref{eq:afbboost}. }
\label{tab:LHC}
\end{table}

The additional faking contributions of $\Wp t$ to the semi-leptonic and dilepton channels is not in conflict with the established data at LHC7. However, recently there are preliminary results from LHC7 that may put stress on the semi-leptonic mode.  Both collaborations report new measurements of the $\sigma(t\bar t)_{lj}$ production cross-section with one $b$-tag, which is relevant for our analysis:
\bea
\sigma(t\bar t)_{lj} & = & 186\pm 10\, {\rm (stat)}^{+21}_{-20}\, {\rm (syst)}\, \pm 6\, {\rm (lumi)}\, {\rm pb}~~{\rm (ATLAS)}\, \\
\sigma(t\bar t)_{lj} & = & 150 \pm 9\, {\rm (stat)}\, \pm 17\, {\rm (syst)}\, \pm 6\, {\rm (lumi)}\, {\rm pb}~~{\rm (CMS)}\,
\eea
The ATLAS measurement \cite{Atlas:ljprelim} is made upon analyzing $35\, {\rm pb}^{-1}$ of data, and the CMS measurement \cite{CMS:ljprelim} is made upon analyzing $36\, {\rm pb}^{-1}$ of data. The ATLAS measurement is well consistent with the $193\, {\rm pb}$ rate that our model point A predicts with $\cos\theta=0.95$. The CMS result is lower and on the surface looks to be a $\sim 2\sigma$ deviation from out prediction.  However, this result is very preliminary and its error is completely systematics dominated. Furthermore, as we have discussed earlier, the details of cuts and acceptances make a significant difference in how much faking contribution there is to the signal, so we are hesitant to make too strong a statement about the applicability of this bound. Nevertheless, if this issue turns out to not mitigate the discrepancy when the details are revealed, and more importantly, if the ATLAS result begins to push more toward the CMS result rather than vice versa, our model will need to predict a lower rate for $\sigma(t\bar t)_{lj}$. This can be achieved straightforwardly by reducing the value of $\cos\theta$, as can be seen by the results of table~\ref{tab:costh}. This does not create conflict with other observables as long as $\cos\theta>0.92$, which is low enough to presently give a prediction of $\sigma(t\bar t)_{lj}\simeq 180\, {\rm pb}$ in line with the CMS result.

\begin{table}
\begin{tabular}{c @{\hspace{1cm}} c @{\hspace{0.5cm}} c @{\hspace{0.5cm}} c}
\hline \hline
$\cos \theta$ (Point A) & $\sigma(t\bar{t})_{\ell j}$ & $\sigma(t\bar{t})_{\ell \ell}$ & $\sigma(tt,\,\bar{t}\bar{t})$ \\
\hline
0.9  & 175 pb & 166 pb & 3.90 pb \\
0.95 & 193 pb & 177 pb & 1.34 pb \\
1.0  & 233 pb & 216 pb & 0 pb \\
\hline \hline
\end{tabular}
\caption{The $\cos \theta$ dependence of two relevant LHC signals (7 TeV) for a mass spectrum corresponding to Point A. Point A is defined with $\cos\theta=0.95$ but other values in the range of $0.92\lesssim \cos\theta \lesssim 0.98$ are also allowed. The inferred inclusive $t\bar{t}$ cross-sections are shown in $\sigma(t\bar{t})_{\ell j,\, \ell \ell}$. Refer to table~\ref{tab:LHC} for more details and corresponding LHC7 data. The inclusive like-sign top pair production (including $tt$, $\bar{t} \bar{t}$ and vector boson decays to like-sign tops) is calculated at LO. Current deduced upper bound of $\sigma(tt,\bar{t}\bar{t})$ from heavy exotic quark searches at LHC7 is about 5 pb at 95$\%$ CL.}
\label{tab:costh}
\end{table}

The observation of like-sign top pairs interestingly turns out to be an important signature for this model even though the cross-section is suppressed by small $\sin \theta$. In the limit $\cos \theta \rightarrow 1$, top-number is preserved and no like-sign top pairs are produced. As discussed earlier, the constraints  from the dijet events already enforce $\theta$ to be relatively small but nonzero. The benchmark point $A$  predicts the total inclusive LO like-sign top pair productions to be 20.6 fb at the Tevatron, and 1.34 pb at LHC7. These inclusive cross-sections include contributions from $qq \rightarrow tt$, $q\bar{q} \rightarrow V V$, $gq \rightarrow tV$ and their charge conjugate processes (where V = $\Wp$, $\Zp$). Current bounds from the Tevatron are more than an order of magnitude weaker than this \cite{LSDL:2007dz}.

The situation for like-sign tops at the LHC is more subtle at the moment, but more promising.  We can extract a relevant bound from LHC7 coming from a like-sign dilepton search (combined with tri-lepton topology) used to search for heavy bottom-like quarks with an integrated luminosity of 34 pb$^{-1}$ \cite{Chatrchyan:2011em}. Their 95\% CL bound on the cross-section of these exotic heavy quarks is about 2-3 pb. Accounting for differences in branching ratios, we estimate a rough 95\% CL limit of 5 pb on our like sign top production to be compared with our prediction of 1.34 pb.     So, while nothing has been seen up to the present, data from the next year may prove relevant.  This is an  interesting conclusion:   while our model completely suppresses the like-sign top quarks in the $\cos \theta \rightarrow 1$ limit,  the like-sign dilepton signal again becomes important once we deviate from this point.  We stress that  $\cos \theta \ne 1$ is necessary.  Without a non-zero $\theta$, we violate the dijet constraints from UA2 and CDF (see Fig. \ref{fig:dijetbounds}), and the faking contribution of the associated $V^{\prime}$ production can become more dangerous. After considering various constraints on the model in previous sections, we found the allowed range of $0.92 \lesssim \cos \theta \lesssim 0.98$ for Point $A$ which is largely set by dijet resonance bounds on $\Wp$ (lower bound) and $\Zp$ (upper bound) as alluded to earlier, see discussion in Sec.~\ref{sec:dijet}.

The values of the like-sign top pair are tabulated for different values of $\cos \theta$ in Table~\ref{tab:costh}. Whichever $\cos \theta$ is realized a signal in some channel is anticipated soon at LHC7.

\begin{figure}
\includegraphics[width=0.48\textwidth]{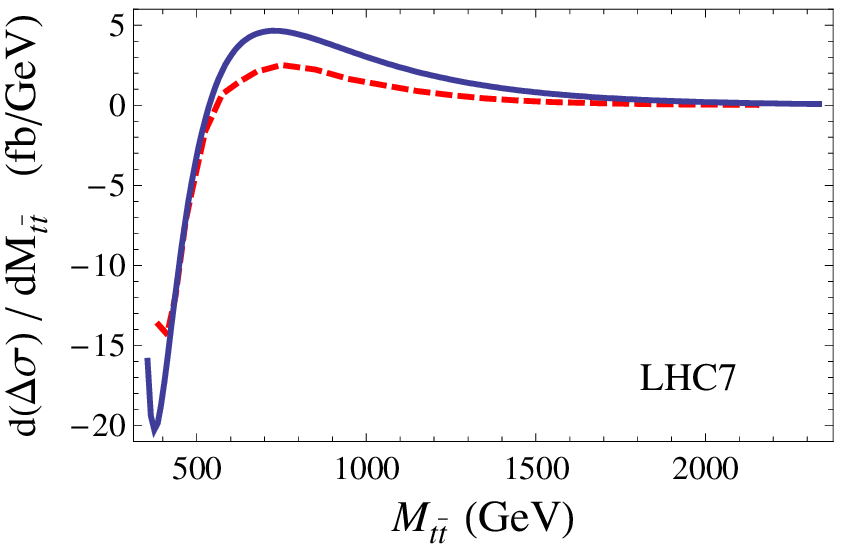}
\includegraphics[width=0.48\textwidth]{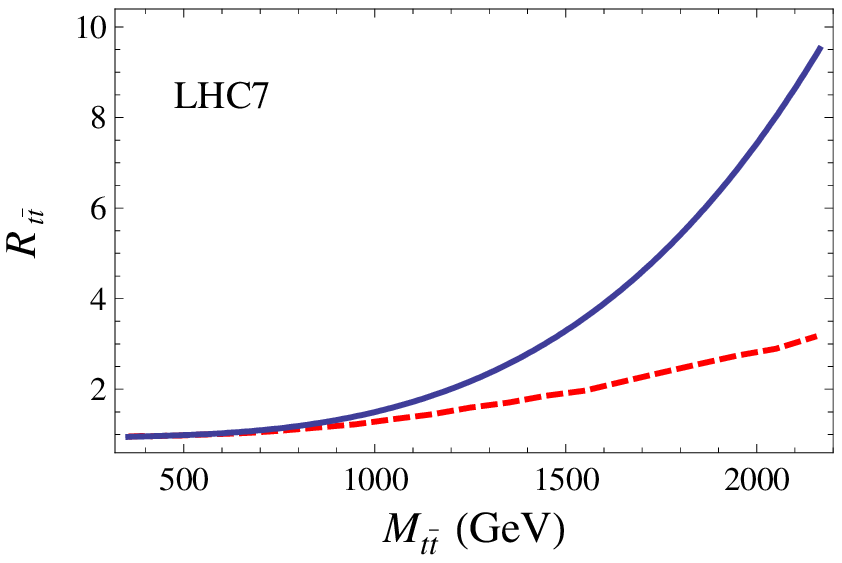}
\caption{ $d( \Delta \sigma )/dM_{t\bar{t}}$ and $R_{t\bar{t}}$ are plotted at LHC7, as defined in Eq.~\ref{eq:mttobs}. The full theoretical distribution (solid line) is subject to acceptance issue as discussed in text. To see the effects of selection cuts, we also calculate distribution by restricting to $y_t,y_{\bar{t}} < 2$ (dashed line). }
\label{fig:mttLHC}
\end{figure}

A large enhancement at high $m_{t\bar{t}}$ is still expected to be present at the LHC as shown in Fig.~\ref{fig:mttLHC}.  See \cite{AguilarSaavedra:2011vw} for related work. Although the $q\bar{q}\to t\bar{t}$ subprocess is less important than $gg\to t\bar t$ at the LHC, the $gg$-initiated subprocess is well suppressed in the high-energy region. Two observables that are sensitive to an enhancement from $q\bar q$ initiated top quark production are defined as
\beq
\frac{ d(\Delta \sigma) }{ dM_{t\bar{t}} } \, \equiv \, \frac{ d\sigma(A) }{ dM_{t\bar{t}} } - \frac{ d\sigma(SM) }{ dM_{t\bar{t}} }, \qquad R_{t\bar{t}} \, \equiv \, \frac{ d\sigma(A)/dM_{t\bar{t}} }{ d\sigma(SM)/dM_{t\bar{t}} }
\label{eq:mttobs} \eeq
and are shown in Fig.~\ref{fig:mttLHC}. However, the differential cross section is subject to the unfolding issue as discussed earlier. To see the important effects of event selection cuts, we also calculate the above two observables by restricting to small top quark rapidities of $y_t,y_{\bar{t}} < 2$. These simple cuts illustrate the acceptance issue well, as forward top quarks have large rapidities. Under these cuts, theoretical distributions are distorted as shown in Fig.~\ref{fig:mttLHC}. It is clear that understanding the acceptance issue at the LHC is very important. Moreover, the high-energy region is expected to be also sensitive to heavy new physics \cite{Delaunay:2011gv}.

To conclusively discover $t$-channel physics and distinguish it from other candidates of  new physics,  the reconstruction of the flavor violating gauge bosons should be attempted. One possibility is to search for the resonance of $\Wp$ through the process $gu \to W^{\prime -} t \to (u\bar{t})\, t$ \cite{Shu:2009xf, Cheung:2011qa, Gresham:2011dg} or through $gu \to W^{\prime -} t \to (u \bar{u}) \, t\,$ if $\cos \theta < 1$ \cite{Jung:2009jz}. However, non-zero $\theta$ in our model reduces the discovery reach a bit. To illustrate this, we calculate relevant cross-sections. For $\cos \theta = 1,\,0.95,\,0.9$, the relevant cross-sections are $\sigma(gu \to \Wp t) \cdot {\rm Br} (\Wp \to tu) \,= \, 58,\, 22,\, 11$ pb at LO.
The later stage of LHC7 may be sensitive to the existence of a light $\Wp$. And lastly, although a light $\Zp$ boson was important to determine plausible benchmark points, it is unlikely to be discovered first since growing QCD backgrounds at low energy wash out the dijet resonance signal.

In summary, the forward-backward asymmetry is a tantalizing signal of new physics at the Tevatron. The non-Abelian symmetry that we have introduced here works well to match the data.  Confirming that it is a correct theory will require careful measurements of the top quark production cross-section and the $t\bar t$ invariant mass distributions at the LHC, as well as like-sign top quark signatures that are made possible by exotic interactions with the new gauge bosons.

\section*{ACKNOWLEDGMENTS}

We thank D.~Amidei, T.~Schwarz, and H.~Murayama for useful discussions.  This work is supported by the DOE Grant \#DE-FG02-95ER40899. The work of AP is supported in part by NSF Career Grant NSF-PHY-0743315.

\appendix
\section{cross-sections and Decay widths}\label{app:xsec}

We list analytic cross-section formula relevant for top pair production. Our notation is that a $q\bar{q} \to t\bar{t}$ diagram denoted by $tX$ implies that a gauge boson $X$ is exchanged in the $t$-channel. Angular differential cross-sections are given by

\beq
\frac{d \sigma_{(sG-tV)}}{d \cos \theta} \= 2 \cdot \frac{\pi \beta}{18 s} \alpha_S \alpha_X \xi_{tV} \cdot \frac{4(u_t^2+sm_t^2) + 2\frac{m_t^2}{m_V^2}(t_t^2+sm_t^2)}{s \cdot t_V}
\eeq
\beq
\frac{d \sigma_{(tV-tV)}}{d \cos \theta} \= \frac{\pi \beta}{8s} \alpha_X^2 \xi_{tV}^2 \cdot \frac{4u_t^2 + \frac{m_t^4}{m_V^4}(t_t^2+4 s m_V^2)}{ t_V^2}
\eeq
\beq
\frac{d \sigma_{(tX-tY)}}{d \cos \theta} \= 2 \cdot \frac{\pi \beta}{8s} \alpha_X^2 \xi_{tX} \xi_{tY} \cdot \frac{4u_t^2 + \frac{m_t^4}{m_X^2 m_Y^2}t_t^2 + 2\frac{m_t^4}{m_X^4} s m_X^2 + 2\frac{m_t^4}{m_Y^4} s m_Y^2}{ t_X \cdot t_Y}
\eeq
\beq
\frac{d \sigma_{(sX-tY)}}{d \cos \theta} \= 2 \cdot \frac{\pi \beta}{24 s} \alpha_X^2 \xi_{sX} \xi_{tY} \cdot \frac{ 4u_t^2 + 2 \frac{m_t^2}{m_Y^2} s m_t^2 }{ s_X \cdot t_Y }
\eeq
\beq
\frac{d \sigma_{(sV-sV)}}{d \cos \theta} \= \frac{\pi \beta}{8 s} \alpha_X^2 \xi_{sV}^2 \cdot \frac{ 4u_t^2 }{ s_V^2 }
\eeq
\beq
\frac{d \sigma_{(sX-sY)}}{d \cos \theta} \= 2 \cdot \frac{\pi \beta}{8 s} \alpha_X^2 \xi_{sX} \xi_{sY} \cdot \frac{ 4u_t^2 }{ s_X \cdot s_Y }
\eeq
where $t_i \equiv t - m_i^2$ (and similarly for $s_i$ and $u_i$), $\beta \equiv \sqrt{(1-4m_t^2/s)}$, and $\xi_{tV}$ is a vertex factor of the $tV$ diagram that can be read from the interaction Lagrangian in Eq.~\ref{interaction}. For example,
\beq
\xi_{t\Wp} \= \frac{1}{2} \,( c^4 + s^4 ), \qquad \xi_{s\Wp} \= \frac{1}{2} \, ( -2 c^2 s^2 )
\eeq
\beq
\xi_{t\Zp} \= \frac{1}{4} \,( 4 c^2 s^2 ), \qquad \xi_{s\Zp} \= \frac{1}{4} \, ( -(c^2 - s^2)^2 )
\eeq
where $c=\cos \theta$ parameterizes the mismatch between the gauge and mass eigenstates of the $(t,u)_R$ doublet.

Decay widths of $\Wp$ and $\Zp$ are given by
\beq
\Gamma( W^{\prime +} \to t \bar{u} ) \= \frac{N_c \alpha_X c^4}{24} M_{\Wp} (1-\xi^2)(2-\xi^2-\xi^4)
\eeq
\beq
\Gamma( W^{\prime +} \to u \bar{t} )_{breaking} \= \frac{N_c \alpha_X s^4}{24} M_{\Wp} (1-\xi^2)(2-\xi^2-\xi^4)
\eeq
\beq
\Gamma( \Zp \to u \bar{t}, t\bar{u} ) \= \frac{N_c \alpha_X \,(8 c^2 s^2)}{48} M_{\Zp} (1-\xi^2)(2-\xi^2-\xi^4)
\eeq
\beq
\Gamma( W^{\prime +} \to u \bar{u} ) \= \frac{N_c \alpha_X c^2 s^2}{12} M_{\Wp}
\eeq
\beq
\Gamma( \Zp \to u \bar{u} ) \= \frac{N_c \alpha_X (s^2-c^2)^2}{24} M_{\Zp}
\eeq
\beq
\Gamma( W^{\prime +} \to t \bar{t} ) \= \frac{N_c \alpha_X c^2 s^2}{12} M_{\Wp} \sqrt{1-4\xi^2}( 1- \xi^2 )
\eeq
\beq
\Gamma( \Zp \to t \bar{t} ) \= \frac{N_c \alpha_X (c^2- s^2)^2}{24} M_{\Zp} \sqrt{1-4\xi^2}( 1- \xi^2 )
\eeq
where $\xi \equiv m_t/m_V$. The subscript ``breaking" is to distinguish the top--number breaking decay mode $W^{\prime +} \to u \bar{t}$ from the top--number preserving mode $W^{\prime +} \to t \bar{u}$.

For model point A, the widths of $\Wp$ and $\Zp$ are given by $\Gamma_{\Wp} / M_{\Wp} \simeq 0.0024$ and $\Gamma_{\Zp} / M_{\Zp} \simeq 0.0073$. Thus resonance search should be sensitive, as discussed in Sec.~\ref{sec:dijet}.

\bibliography{topSU2}
\bibliographystyle{apsrev}

\end{document}